\documentclass[aip,jcp,amsmath,amssymb,reprint]{revtex4-1}

\usepackage{graphicx}
\usepackage{amsmath}
\usepackage{color}

\begin{document}

\title{{ Analysis of fluctuations in the first return times of random walks on regular branched networks}}

\author{Junhao Peng}
\affiliation{
  School of Math and Information Science, Guangzhou University,
  Guangzhou 510006, China.}
\affiliation{
  Key Laboratory of Mathematics and Interdisciplinary Sciences of
  Guangdong Higher Education Institutes, Guangzhou University, Guangzhou
  510006, China.}
   \affiliation{
  Center for Polymer Studies and Department of Physics, Boston
  University, Boston, MA 02215, USA}
\author{Guoai Xu}
\affiliation{
School of Cyberspace Security, Beijing University of Posts and
Telecommunications, Beijing 100876, China}

\author{Renxiang Shao}
\affiliation{
  School of Math and Information Science, Guangzhou University,
  Guangzhou 510006, China.}
\affiliation{
  Key Laboratory of Mathematics and Interdisciplinary Sciences of
  Guangdong Higher Education Institutes, Guangzhou University, Guangzhou
  510006, China.}

\author{Lin Chen}
\affiliation{
  School of Management, Northwestern Polytechnical University, Xian
  710072, China.}
   \affiliation{
  Center for Polymer Studies and Department of Physics, Boston
  University, Boston, MA 02215, USA}

\author{H. Eugene Stanley}
\affiliation{
  Center for Polymer Studies and Department of Physics, Boston
  University, Boston, MA 02215, USA}

\begin{abstract}

\noindent
The first return time (FRT) is the time it takes a random walker to
first return to its original site, and the global first passage time
(GFPT) is the first passage time for a random walker to move from a
randomly selected site to a given site.
{  We find that in finite networks the variance of FRT, Var(FRT), can be expressed Var(FRT)~$=2\langle$FRT$
\rangle \langle $GFPT$ \rangle -\langle $FRT$ \rangle^2-\langle $FRT$
\rangle$, where $\langle \cdot \rangle$ is the mean of the random
variable. Therefore  a method of calculating the variance of FRT on
general finite networks is presented. }
We then calculate Var(FRT) and analyze the
fluctuation of FRT on regular branched networks (i.e., Cayley tree) by using Var(FRT)
and its variant as the metric.
{  We find that the results differ from
those in such other networks as}
 Sierpinski gaskets, Vicsek fractals,
T-graphs, pseudofractal scale-free webs, ($u,v$) flowers, and fractal
and non-fractal scale-free trees.

\end{abstract}


\maketitle

\section{Introduction}
\label{intro}

{  The first return time (FRT), an interesting quantity in the random walk
literature, is the time it takes a random walker to first return to its
original site \cite{Redner07, Metzler-2014}.}
 It is a key indicator of
how quickly information, mass, or energy returns back to its original
site in a given system. It can also be used to model the time intervals
between two successive extreme events, such as traffic jams, floods,
earthquakes, and droughts \cite{BunKr05,LeadLin83, ReThRe97,
  KondVa06,BatGer02}. Studies of FRT help in the control and forecasting
of extreme events \cite{KishSant11, ChenHu14}. In recent years much
effort has been devoted to the study of the statistic properties
\cite{EiKaBu07, Nicolis07, MoDa09, LiuJi09, HadLue02} and the
probability distribution \cite{MuSu10,LowMast00,MaKo04, SaKa08,PaPen11,
  Bunde01, Bunde05, IzCa06, Olla07,Ch11} of the FRT in different
systems. {  A wide variety of experimental records show that return
probabilities tend to  exponentially decay \cite{SaKa08,PaPen11, Bunde01,
  Bunde05}. Other findings include the discovery of an interplay between
Gaussian decay and exponential decay in the return probabilities of
quantum systems with strongly interacting particles \cite{IzCa06}, and
the power-law decay in time of the return probabilities in some
stochastic processes of extreme events and of random walks on scale-free
trees \cite{Olla07,Ch11}.

Statistically, in addition to its probability distribution, the mean and
variance of any random variable $T$ are also useful characterization
tools.  The mean $\langle T \rangle$ is the expected average outcome
over many observations and can be used for estimating $T$. The variance
Var$(T)$  is the expectation of the squared deviation of $T$  from its mean
and can be used for measuring the amplitude of the fluctuation of $T$. The reduced moment
of $T$, $R(T)=\frac{\sqrt{\rm Var(T)}}{\langle T \rangle}$
\cite{HaRo08}, is a metric for the relative amplitude
of the fluctuation of T derived by a comparison with its
mean,
  and it can be used to evaluate
whether $\langle T \rangle$ is a good estimate of $T$.} The greater the
reduced moment, the less accurate the estimate provided by the
mean. If $R(T)\rightarrow\infty$, as network size
$N\rightarrow\infty$, the standard deviation  $\sqrt{\rm Var(T)} \gg
\langle T\rangle$. Then we can affirm  that the fluctuation of
$T$ is huge in the network with large size, and that $\langle T \rangle$ is not a reliable estimate of
$T$.

For a discrete random walk on a finite network, the mean FRT can be
directly calculated from the stationary distribution. For an arbitrary
site $u$, $\langle $FRT$\rangle=2E/d_u$, where $E$ is the total number
of network edges and $d_u$ is the degree of site $u$
\cite{LO93}. However the variance Var(FRT) and the reduced moment of FRT
are not easy obtained, and the fluctuation of FRT is unclear. Whether
$\langle $FRT$\rangle$ is a good estimate of FRT is also unclear.

 {  Research shows, the second moment of FRT is closely connected to the frst moment of global first-passage
time (GFPT), which is the first-passage time from a randomly selected
site to a given site \cite{Tejedor09}. We find that in general finite
networks Var(FRT)~$=2\langle $FRT$ \rangle \langle $GFPT$ \rangle
-\langle $FRT$ \rangle^2-\langle $FRT$ \rangle$.We can also derive
Var(FRT) and $R$(FRT) because $\langle $GFPT$ \rangle$ has been
extensively studied and can be exactly derived on a number of different
networks \cite{ChPe13, MeyChe11, CondaBe07, BenVo14, KiCarHa08}. Thus we
can also analyze the fluctuation of FRT and determine when $\langle
$FRT$\rangle$ is a good estimate of FRT.}

As an example, we analyze  the fluctuation of FRT on Cayley trees by using $R$(FRT) as the metric. We  obtain the exact results for Var(FRT) and $R$(FRT), and present their scalings with network size $N$.  {  We use Cayley trees for the following  reasons.}  Cayley trees, also known as dendrimers,
are an important kind of polymer networks. Random walk on Cayley trees \cite{ChCa99, MuBi06, FuDo12} has many
applications, including light harvesting \cite{BarKla97, BarKla98,
  Bentz03, Bentz06} and energy or exciton transport \cite{Sokolov97,
  Blumen81}. First passage problems in Cayley trees have received
extensive study and the $\langle GFPT \rangle$ to an arbitrary target
node has been determined \cite{WuLiZhCh12, LiZh13}. In contrast to
other networks, the $R$(FRT) of Cayley trees differs when the network size
$N\rightarrow\infty$. We find that $R($FRT$)\rightarrow\infty$ on many
networks, including Sierpinski gaskets \cite{BeTuKo10, Weber_Klafter_10,
  Wu_Zhang_10}, Vicsek fractals \cite{BlumenFer04, LiZh13, Peng14d},
T-graphs \cite{Agliar08, ZhangLinWu09, Peng14b}, pseudofractal
scale-free webs \cite{ZhQiZh09, PengAgliariZhang15}, ($u,v$) flowers
\cite{ZhangXie09, MeAgBeVo12, Hwang10, Peng17}, and fractal and
non-fractal scale-free trees \cite{ZhLi11, Peng14a, Peng14c, CoMi10,
  LinWuZhang10}. Thus the fluctuation of FRT in these networks is huge
and the ${\langle FRT \rangle}$ is not a reliable FRT estimate. For
dendrimers, however, $R(FRT)\rightarrow const$ for most cases. Thus the FRT fluctuation is relatively small and
the $\langle FRT \rangle$ is an acceptable FRT estimate.

This paper is structured as follows. Section~\ref{sec:TF} presents the
network structure of the Cayley trees. Section~\ref{sec:Rel_GFPT_FRT}
presents and proves the exact relation between Var(FRT) and $\langle
$GFPT$ \rangle$ on general finite
networks. Section~\ref{sec:Rel_GFPT_FRT} also briefly introduces the
asymptotic results of $R$(FRT) on some networks, which shows that
$R($FRT$)\rightarrow\infty$ and $N\rightarrow\infty$.
Section~\ref{sec:VarienceFRT} presents the explicit results of Var(FRT)
and $R$(FRT), together with  fluctuation analysis of FRT on Cayley trees.
Finally, Sec.~\ref{sec:Conclusion} is left for conclusions and discussions. Technicalities on calculations are collected in the Appendices.

\section{Network structure and properties}
\label{sec:TF}

\begin{figure}
\begin{center}
\includegraphics[scale=0.85]{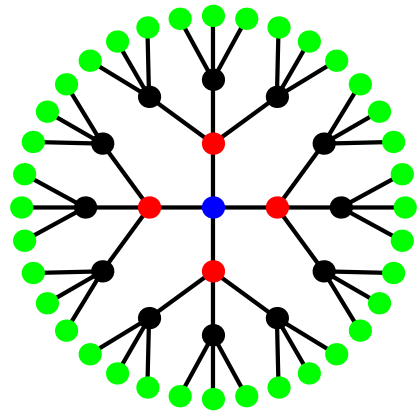}
\caption{Structure of the a particular Cayley tree $C_{4,3}$. Nodes
  colored with red constitute the first shell of $C_{4,3}$; Nodes
  colored with black constitute the second shell of $C_{4,3}$; Nodes
  colored with green constitute the third shell of $C_{4,3}$.}
\label{fig:1}
\end{center}
\end{figure}

 {  The Cayley tree is rooted, and all other nodes are arranged in shells
around its root node \cite{Ostilli12}. It is a regular branched network, where each non-terminal node is connected to $m$  neighbours, and $m$ is called the order of the Cayley tree.
 Here $C_{m,g}(m\geq 3, g\geq 0)$
is a Cayley tree of order $m$ with $g$ shells. Beginning with the root
node,} $m$ new nodes are introduced and linked to the root by $m$
edges. This first set of $m$ nodes constitutes the first shell of
$C_{m,g}$. We then obtain the shell $i$ $(2 \leq i \leq g )$ of
$C_{m,g}$. We add and link $m-1$ new nodes to each node of shell
$(i-1)$. The set of these new nodes constitutes shell $i$ of
$C_{m,g}$. FIG.~\ref{fig:1} shows the construction of a specific
Cayley tree $C_{4,3}$.

Using the construction, one can find all nodes in the same shell are
equivalent. The nodes in the outermost shell have a degree $d_g=1$, and
all other nodes have a degree $d_i=m$ $(i=0,1, \cdots, g-1)$. We also
find that the number of nodes of $(i=1, 2,\cdots, g )$ shell $i$ is
$N_{i}=m(m-1)^{i-1}$. Thus for $C_{m,g}$ the total number of nodes is
\begin{equation}\label{NodeC}
N=1 + \sum_{i=1}^{g}N_{i}=\frac{m(m-1)^g-2}{m-2}\,,
\end{equation}
and the total number of edges in $C_{m,g}$ is
\begin{equation}\label{EegeC}
E=N-1=\frac{m(m-1)^g-m }{m-2}\,.
\end{equation}
 {  Although Cayley trees are obviously self-similar, their fractal
dimension is infinite, and they are thus nonfractal.}

\section{Relation between Var(FRT) and $\langle $GFPT$ \rangle$ on
  general finite networks and results of $R$(FRT) on some networks}
\label{sec:Rel_GFPT_FRT}

{ In this section, we present and prove  the general relation between the variance of FRT and the
mean global first-passage time on general finite networks. Our
derivations are based on the relation between their probability generating
functions. To briefly review the definition probability-generating
function (see e.g., \cite{Gut05}), we designate $p_k$ ($k=0,1,2,\cdots$)
the probability mass function of a discrete random variable $T$ that
takes values of non-negative integers $\{0,1, ...\}$, and we define the
related probability-generating function $\Phi_{T}(z)$ of $p_k$,}
\begin{equation}\label{Def_PGF}
\Phi_{T}(z)=\sum_{k=0}^{+\infty}z^k p_k.
\end{equation}


Now we  introduce  the probability distribution of  GFPT and FRT, and then define the probability generating functions of them.  Before proceeding, we must clarify that,  when evaluating the GFPT, the starting site is selected by mimicking the  steady state, namely the probability that a node $u$ is selected as starting site is $d_u/(2E)$. 

 {  Here $P_{v\rightarrow u}(k)$ ($k=0,1,2,\cdots$) is the probability
distribution of the first passage time (FPT) from node $v$ to node
$u$. Thus $P_{u \rightarrow u}(k)$ ($k=0,1,2,\cdots$) is the
probability distribution of FRT when the target is located at node
$u$. The probability distribution of the GFPT to the target node $u$,
denoted as $P_{u}(k)$ ($k=0,1,2,\cdots$), is defined}
{
\begin{equation}\label{Def_GFPP}
  P_{u}(k)=\sum_{v}\frac{d_v}{2E}P_{v\rightarrow u}(k),
\end{equation}
}
where the sum runs over all the nodes in the network.

We denote $\Phi_{\rm FRT}(z)$ and $\Phi_{\rm GFPT}(z)$ the
probability-generating functions of the FRT and GFPT for node $u$,
respectively. Both have a close connection with the probability
generating function of the return time (i.e., how long it takes the
walker to return to its origin, not necessarily for the first time),
whose generating function is $\Phi_{\rm RT}(z)$. Note that
\cite{HvKa12},
\begin{equation}\label{R_GFPT_RT_g}
   \Phi_{\rm GFPT}(z)=\frac{ z}{1-z}\times\frac{d_u}{2E}\times\frac{1}{
     \Phi_{\rm RT}(z)},
\end{equation}
and
\begin{equation}\label{R_FRT_RT_g}
 \Phi_{\rm FRT}(z)=1-\frac{1}{\Phi_{\rm RT}(z)}.
\end{equation}
Equation~(\ref{R_GFPT_RT_g})) can now be rewritten
\begin{equation}\label{RT_GFPT}
   \frac{1}{ \Phi_{\rm RT}(z)}=\frac{1-z}{
     z}\times\frac{2E}{d_u}\times\Phi_{\rm GFPT}(z).
\end{equation}
Plugging the expression for $ \frac{1}{ \Phi_{\rm RT}(z)}$ into
Eq.~(\ref{R_FRT_RT_g}), we get
\begin{equation}\label{R_FRT_GFPT}
\Phi_{\rm FRT}(z)=1-\frac{1-z}{ z}\times\frac{2E}{d_u}\times\Phi_{\rm
  GFPT}(z).
\end{equation}
Taking the first derivative on both sides of Eq.~(\ref{R_FRT_GFPT}) and
setting $z=1$, we obtain the mean FRT,
\begin{eqnarray}\label{First_m_FRT}
\langle FRT \rangle=\left. \frac{d}{d z}\Phi_{\rm
  FRT}(z)\right|_{z=1}=\frac{2E}{d_u}.
\end{eqnarray}
Taking the second order derivative on both sides of
Eq.~(\ref{R_FRT_GFPT}) and setting $z=1$, we obtain
\begin{eqnarray}\label{D2_FRT}
      \left. \frac{d^2}{d z^2}\Phi_{\rm FRT}(z)\right|_{z=1}
      &=&\frac{2E}{d_u}\left\{2\left. \frac{d}{d z}\Phi_{\rm
        GFPT}(z)\right|_{z=1} -2\right\} \nonumber\\
      &=&2\langle FRT  \rangle \langle GFPT  \rangle -2\langle FRT
      \rangle.
\end{eqnarray}
We thus get the variance
\begin{eqnarray}\label{R_Second_m_FRT}
   Var(FRT)&=&\left. \frac{d^2}{d z^2}\Phi_{\rm
     FRT}(z)\right|_{z=1}+\left. \frac{d}{d z}\Phi_{\rm
     FRT}(z)\right|_{z=1}-\langle FRT  \rangle^2 \nonumber\\
     &=&2\langle FRT  \rangle \langle GFPT  \rangle -\langle FRT
   \rangle^2-\langle FRT  \rangle,
\end{eqnarray}
and the reduced moment
\begin{eqnarray}\label{Reduce_m_FRT}
    R(FRT)=\frac{\sqrt{Var(FRT)}}{ \langle FRT\rangle} \approx
    \sqrt{\frac{2\langle GFPT  \rangle}{\langle FRT\rangle}-1}.
\end{eqnarray}

 {  Both $\langle GFPT \rangle$ and $\langle FRT\rangle$ increase with the
increase of network size $N$, and the order in which $\langle GFPT
\rangle$ increases is no less than that of $\langle FRT\rangle$. If the
order that $\langle GFPT \rangle$ increases is greater than that of
$\langle FRT\rangle$, $R(FRT)\rightarrow\infty$ as
$N\rightarrow\infty$. However if the order that $\langle FRT\rangle$
increases is the same as that of $\langle GFPT \rangle$,
$R(FRT)\rightarrow const$ as $N\rightarrow\infty$.}

If $\langle GFPT\rangle$ has been obtained on a network, $R(FRT)$ can
also be obtained on that network.  For example, on classical and dual
Sierpinski gaskets embedded in $d$-dimensional ($d\geq 2$) Euclidian
spaces, $\langle FRT\rangle \sim N$ and $\langle GFPT\rangle \sim
N^{2/d_s}$, where $d_s=\frac{2ln(d+1)}{ln(d+3)}$ \cite{Weber_Klafter_10,
  Wu_Zhang_10}.  Thus $R(FRT)\sim
N^{\frac{ln(d+3)}{2ln(d+1)}-\frac{1}{2}}$ and
\begin{eqnarray}\label{SCa_Reduce_m_FRT}
    R(FRT)\rightarrow\infty,
\end{eqnarray}
 {  as $N\rightarrow\infty$.  We find that Eq.~(\ref{SCa_Reduce_m_FRT})
also holds on many other networks}, such as Vicsek fractals, T-graph,
pseudofractal scale-free webs, ($u,v$) flowers, and fractal and
non-fractal scale-free trees.  {  Although $\langle FRT\rangle$ is easy to
obtain in these networks,} it is not a reliable estimate of FRT because
the fluctuation of FRT is huge.

\section{ Fluctuation analysis of first return time on  Cayley
  trees} \label{sec:VarienceFRT}

We now calculate the variance, the reduced moment of FRT, and then
analyze the fluctuation of FRT on  Cayley trees. Note that the target
location strongly affects Var(FRT) and $R(FRT)$.  We calculate Var(FRT)
and $R(FRT)$ when the target is located at an arbitrary node on  Cayley
trees. We obtain exact results for Var(FRT) and $R(FRT)$ and present
their scalings with network size.  The derivation presented here is
based on the relation between Var(FRT) and $\langle GFPT \rangle$
expressed in Eq.~(\ref{R_Second_m_FRT}). We first thus derive the mean
GFPT \cite{Notes1} to an arbitrary node on $C_{m,g}$. We then obtain
Var(FRT) and $R$(FRT) from Eqs.~(\ref{R_Second_m_FRT}) and
(\ref{Reduce_m_FRT}). Because the calculation is lengthy, we here
summerize the the derivation and the final results and present the
detailed derivation in the Appendix
\ref{APP:MGFPTg}--\ref{APP:MGFPT}.

\subsection{Mean  GFPT and the variance of FRT while the target site is located at arbitrary  node  on Cayley trees} \label{sec:Moments}

Here $\Omega$ is the node set of the Cayley tree $C_{m,g}$, and we
define
\begin{equation}
W_v=\sum_{u\in \Omega}\pi(u)L_{uv},
\label{WV}
\end{equation}
and
\begin{equation}
\Sigma=\sum_{u\in \Omega}\pi(u)W_u,
\label{WS}
\end{equation}
where $L_{uv}$ is the shortest path length from node $u$ to $v$, and
$\pi(u)=\frac{d_u}{2E}$. Using the relation between the mean first
passage time and the effective resistance, if the target site is fixed
at node $y$ ($y \in \Omega$) we find the mean GFPT to node $y$ to be
\begin{eqnarray} \label{MGFPT}
\langle GFPT_y\rangle&=&E(2W_y-\Sigma)+1.
\end{eqnarray}
We supply the detailed derivation in Appendix \ref{APP:MGFPTg}.

Note that the target location strongly affects the moments of GFPT and
FRT, and that all nodes in the same shell of $C_{m,g}$ are
equivalent. 
Here $GFPT_i$, $FRT_i$ $(i=0, 1, 2,\cdots, g )$ are the GFPT
and FRT, respectively, and the target site is located in shell $i$ of
the Cayley tree $C_{m,g}$. Note that we here regard the node in shell
$0$ to be the root of the tree.
 Calculating $W_v$ for any node $v$ and
$\Sigma$ and plugging their expressions into Eq.~(\ref{MGFPT}), we
obtain the mean GFPT to the root,
\begin{eqnarray}
\langle
GFPT_0\rangle&=&\frac{1}{2E}\left[(m-1)^{2g}\frac{4m(m-1)}{(m-2)^3}\right.\nonumber
  \\
                 & &-(m-1)^{g}\frac{m(4gm-3m+6)}{(m-2)^2}\nonumber \\
                 & &\left.-\frac{m(3m^2-8m+8)}{(m-2)^3} \right],
                \label{MGFPT0}
\end{eqnarray}
and the mean GFPT to nodes in  shell $i$  $(i=1, 2,\cdots, g )$ of
$C_{m,g}$,
\begin{eqnarray}
\langle GFPT_i\rangle&=&(m-1)^{g}\frac{2m(m-2)i-4(m-1)}{(m-2)^2}\nonumber \\
                 & &+\frac{4(m-1)^{g-i+1}}{(m-2)^2}+\langle GFPT_0\rangle.  \label{MGFPTi}
\end{eqnarray}
We supply a detailed derivation of Eqs.~(\ref{MGFPT0}) and
(\ref{MGFPTi}) in Appendix \ref{APP:MGFPT}.  Inserting the expressions
for the mean GFPT and mean FRT into Eqs.~(\ref{R_Second_m_FRT}), we
obtain the variance of FRT for $i=0,1, 2,\cdots, g-1 $,

\begin{eqnarray} \label{Var_FRTi}
     Var(FRT_i)&=&(m-1)^{2g}\frac{8m(m-2)i-12m+16}{(m-2)^3}\nonumber \\
         &-&(m-1)^{g}[\frac{8gm-4m+ 8mi}{(m-2)^2}+\frac{16(m-1)}{(m-2)^3}]\nonumber \\
         &+&\frac{16(m-1)^{2g-i+1}}{(m-2)^3}-\frac{16(m-1)^{g-i+1}}{(m-2)^3}\nonumber \\
        & &-\frac{4m^2-4m}{(m-2)^3},
\end{eqnarray}
and for $i=g$,
\begin{eqnarray}
     Var(FRT_g)&=&(m-1)^{2g}\frac{8m^2(m-2)g-4m(m^2-2)}{(m-2)^3}\nonumber \\
         &&-(m-1)^{g}\frac{4m(8gm-4gm^2+3m^2-4)}{(m-2)^3}\nonumber \\
           & &-\frac{8m^3-8m}{(m-2)^3}. \label{Var_FRTg}
\end{eqnarray}
Then the reduced moments of FRT can be exactly determined using
Eq.~(\ref{Reduce_m_FRT}).

\subsection{Scalings}

Using the results found in the previous subsections we derive their
scalings with network size $N$. Note that $N=\frac{m(m-1)^g-2}{m-2}\sim
(m-1)^g$(see Sec.~\ref{sec:TF}). We get $g \sim ln(N)$, $E=N-1\sim N$,
and $\langle FRT_i\rangle\sim N$ for any $i$. We further reshuffle
Eqs.~(\ref{MGFPT0}), (\ref{MGFPTi}), (\ref{Var_FRTi}), and
(\ref{Var_FRTg}) and get for $i=0,1,\cdots ,g$
\begin{eqnarray} \label{S_GFPTi}
\langle GFPT_i\rangle \sim (i+1)N,
\end{eqnarray}
and
\begin{eqnarray} \label{S_Vari}
   Var(FRT_i) \sim (i+1)N^2.
\end{eqnarray}
If we now set $i=g$ in Eqs.~(\ref{S_GFPTi}) and (\ref{S_Vari}) we obtain
\begin{eqnarray} \label{S_GFPTg}
\langle GFPT_g\rangle \sim Nln(N),
\end{eqnarray}
and
\begin{eqnarray} \label{S_Varg}
   Var(FRT_g) \sim N^2ln(N).
\end{eqnarray}
Inserting the expressions for $Var(FRT_i)$ and $\langle FRT_i\rangle$
into Eq.~(\ref{Reduce_m_FRT}), we obtain the reduced moment of FRT and
find that in the large size limit (i.e., when $N\rightarrow \infty$), for
$i=0,1,\cdots ,g-1$,
\begin{eqnarray} \label{S_Ri}
  R(FRT_i)\rightarrow \sqrt{2mi-\frac{3m-4}{m-2}+\frac{4(m-1)^{1-i}}{m-2}},
\end{eqnarray}
and
\begin{eqnarray} \label{S_Rg}
R(FRT_g)\approx \sqrt{4g-\frac{m^2+2m-4}{m(m-2)}}\rightarrow \infty.
\end{eqnarray}
Results show that in the large size limit,
$R(FRT_i)$ increases as $i$ increases, which implies that the farther
the distance between target and root, the greater the fluctuation of
FRT.  If the target site is fixed at the root (i.e. $i=0$), $R(FRT_i)$
reach its minimum
\begin{eqnarray} \label{S_R0}
R(FRT_0)\rightarrow \sqrt{\frac{m}{m-2}}.
\end{eqnarray}
If the target site is fixed at shell $i$ (i.e., $i$ does not increase as
$N$ increases), $R(FRT_i)\rightarrow const$. Here the fluctuation of FRT
is small and $\langle FRT\rangle$ can be used to estimate FRT. If $i$
increases with the network size $N$, e.g.,  the target is located at
the outermost shell (i.e. $i=g\sim ln(N)$), $i \rightarrow \infty$ as
$N\rightarrow \infty$. Thus $R(FRT_i)\rightarrow \infty$. Here the
fluctuation of FRT is huge and $\langle FRT\rangle$ is not a reliable
estimate of FRT.

\section{Conclusions} \label{sec:Conclusion}

{ We have found the exact relation between Var(FRT) and $\langle GFPT
\rangle$ in a general finite network. We thus can determine the exact
variance Var(FRT) and reduced moment $R(FRT)$ because $\langle GFPT
\rangle$ has been widely studied and measured on many different
networks. We use the reduced moment to measure and evaluate the
fluctuation of a random variable and to determine whether the mean of a
random variable is a good estimate of the random variable. The greater
the reduced moment, the worse the estimate provided by the mean.}

In our research we find that in the large size limit (i.e., when
$N\rightarrow\infty$), $R(FRT)\rightarrow\infty$, which indicates that
FRT has a huge fluctuation and that $\langle FRT \rangle$ is not a
reliable estimate of FRT in most networks we studied. However for random walks on Cayley trees, in most cases,  $R(FRT)\rightarrow const$. 

We also find that target location strongly affects  FRT fluctuation on
Cayley trees. Results show that the farther the distance between target and root, the
greater the FRT fluctuations. $R(FRT)$ reaches its minimum when the
target is located at the root of the tree, and $R(FRT)$ reaches its
maximum when the target is located at the outermost shell of the
tree. Results also show that when the target site is fixed at shell $i$
(i.e., $i$ does not increase as $N$ increases), $R(FRT_i)\rightarrow
const$. Here the fluctuation of FRT is small and $\langle FRT\rangle$
can be used to estimate FRT. When $i$ increases with network size $N$
(e.g., i=g), $R(FRT_i)\rightarrow \infty$. Here the fluctuation of FRT
is huge and $\langle FRT\rangle$ is not a reliable estimate of FRT.

\acknowledgments { The Guangzhou University School of Math and Information Science is supported by China Scholarship
  Council (Grant No. 201708440148),  the special innovation project of colleges and universities in
  Guangdong(Grant No. 2017KTSCX140), and by the National Key R\&D Program of China (Grant No. 2018YFB0803700).
   The Northwestern Polytechnical Universitwas  School of Management is supported by the National Social Science Fund of   China (17BGL143), Humanities and Social Science Talent Plan of Shaanxi
  and China Postdoctoral Science Foundation (2016M590973).
  The Boston  University Center for Polymer Studies is supported by NSF Grants
  PHY-1505000, CMMI-1125290, and CHE-1213217, and by DTRA Grant  HDTRA1-14-1-0017.  }

\appendix


\section{Derivation of Eq.~\ref{MGFPT}}
\label{APP:MGFPTg}

We denote $\Omega$ the node set of any graph $G$. For any two nodes $x$
and $y$ of graph $G$, $F(x,y)$ is the mean FPT from $x$ to $y$. Therefore $F(y,y)$ is just the first return time for node $y$. For any different two nodes $x$ and $y$,  the sum
$$k(x,y)=F(x,y)+F(y,x)$$ is the commute time, and the mean FPT can be
expressed in terms of commute times \cite{Te91}
\begin{equation}
F(x,y)=\frac{1}{2}\left(k(x,y)+\sum_{u\in \Omega}\pi(u)[k(y,u)-k(x,u)]
\right),
\label{FXY}
\end{equation}
where $\pi(u)=\frac{d_u}{2E}$ is the stationary distribution for random
walks on the $G$, $E$ is the total numbers of edges of graph $G$, and
$d_u$ is the degree of node $u$.

We treat these systems as electrical networks, consider each edge a unit
resistor, and denote $\mathfrak{R}_{xy}$ the effective resistance
between nodes $x$ and $y$. Prior research \cite{Te91} indicates that
\begin{equation}
k(x,y)=2E\mathfrak{R}_{xy}.
\label{KR}
\end{equation}
If graph $G$ is a tree, the effective resistance between any two nodes
is the shortest path length between the two nodes. Hence
\begin{equation}
 \mathfrak{R}_{xy}=L_{xy},
\end{equation}
where $L_{xy}$ is the shortest path length between node $x$ to node $y$.
Thus
\begin{equation}
k(x,y)=2EL_{xy}.
\label{KL}
\end{equation}
Substituting $k(x,y)$ in the right side of Eq.~(\ref{KL}), in
Eq.~(\ref{FXY}) the mean FPT from $x$ to $y$ can be rewritten
\begin{eqnarray}
F(x,y) &=&E(L_{xy}+W_y-W_x).
\label{FXYL}
\end{eqnarray}
Thus the mean GFPT to $y$ can be written
%
{ \begin{eqnarray} \label{MTT_a}
\langle GFPT_y \rangle&=&\sum_{x\in \Omega}\pi(x)F(x,y) \nonumber  \\
&=&\pi(y)F(y,y)+\sum_{x \neq y}\pi(x)F(x,y) \nonumber  \\
&=&1+\sum_{x \neq y}\pi(x)E(L_{xy}+W_y-W_x) \nonumber  \\
&=&1+E\sum_{x \neq y}\pi(x)L_{xy}+E\sum_{x \neq y}\pi(x)W_y \nonumber  \\
 & &-E\sum_{x \neq y}\pi(x)W_x) \nonumber  \\
 &=&1+EW_y+E(1-\pi(y))W_y \nonumber  \\
 & &-E\sum_{x \neq y}\pi(x)W_x) \nonumber  \\
&=&E(2W_y-\Sigma)+1.
\end{eqnarray}}

\section{Derivation of Eqs.~(\ref{MGFPT0}) and
  (\ref{MGFPTi})}\label{APP:MGFPT}

We here derive $\langle GFPT_{i} \rangle$ for any $(i=0, 1, 2,\cdots, g
)$. To calculate $\langle GFPT_{i} \rangle$, we assume the target is
located at node $v_i$ in shell $i$ of $C_{m,g}$. Thus $\langle GFPT_{i}
\rangle$ can also be denoted $\langle GFPT_{v_i} \rangle$. Using
Eq.~(\ref{MGFPT}), we calculate $W_{v_i}$ and $\Sigma$ defined in
Eqs.~(\ref{WV}) and (\ref{WS}).

Calculating the shortest path length between any two nodes in a Cayley
tree is straightforward, e.g., the shortest path length between
arbitrary node $u$ in shell $i$ is $(i=1,2,\cdots,g)$ and the root $v_0$
is $L_{uv_0}=i$. Thus for root node $v_0$,

\begin{eqnarray}
  W_{v_0}&\equiv&\sum_{u\in \Omega}\pi(u)L_{uv_0} \nonumber \\
  &=&\frac{1}{2E}\left\{ \sum_{i=1}^{g-1}[m^2
    i(m-1)^{i-1}]+mg(m-1)^{g-1} \right\}\nonumber \\
  &=& \frac{m(m - 1)^g( 2gm-4g- m)+m^2}{2E\times(m - 2)^2} .
\label{W0}
\end{eqnarray}
{ For arbitrary node $v_i$  in shell $i$ ($i=1, 2,\cdots, g$), we can find its parents node in shell $i-1$ and let $v_{i-1}$ denotes the parents node of $v_i$ and $\Omega_i$ denotes the nodes set of the subtree whose root  is $v_i$.
  We find that
 \begin{equation}
\label{Luvi1}
  L_{uv_i}-L_{uv_{i-1}}=\left\{ \begin{array}{ll}  -1 & u \in \Omega_i  \\1  & \textrm{otherwise} \end{array} \right..\nonumber
\end{equation}
Hence,
\begin{eqnarray}
  &&W_{v_i}-W_{v_{i-1}} \nonumber \\
  &=&\sum_{u\in \Omega}\pi(u)(L_{uv_i}-L_{uv_{i-1}}) \nonumber \\
  &=&\sum_{u\in \Omega}\pi(u)-2\sum_{u\in \Omega_i}\pi(u) \nonumber \\
  &=&1-\frac{1}{E}\left\{ \sum_{k=i}^{g-1}m(m-1)^{k-i}+(m-1)^{g-i} \right\}\nonumber \\
  &=& 1-\frac{1}{E}\left\{ \frac{2}{m-2}(m-1)^{g-i+1}-\frac{m}{m-2} \right\}. \nonumber
\label{DWI}
\end{eqnarray}
Thus,
\begin{eqnarray}
  &&W_{v_i}-W_{v_{0}} \nonumber \\
  &=&\sum_{k=1}^i(W_{v_k}-W_{v_{k-1}}) \nonumber \\
  &=& i-\frac{1}{E}\left\{ \sum_{k=1}^i\frac{2}{m-2}(m-1)^{g-k+1}-\frac{mi}{m-2} \right\}. \nonumber \\
  &=& i-\frac{2(m-1)^{g+1}-2(m-1)^{g-i+1}-mi(m-2)}{(m-2)^2E} \nonumber \\
  &=& \frac{mi(m-2)(m - 1)^g-2(m - 1)^{g + 1}}{E\times(m - 2)^2}\nonumber \\
  &+& \frac{ 2(m - 1)^{g -i+ 1} }{E\times(m - 2)^2}.
\label{DWI0}
\end{eqnarray}
Therefore, for any $i=1, 2,\cdots, g$,
\begin{eqnarray}
&&\langle GFPT_i\rangle-\langle GFPT_0\rangle=2E(W_{v_i}-W_{v_0})\nonumber \\
      &=&(m-1)^{g}\frac{2m(m-2)i-4(m-1)}{(m-2)^2}\nonumber \\
                 & &+\frac{4(m-1)^{g-i+1}}{(m-2)^2}, \label{MGFPTid}
\end{eqnarray}
and Eq.~(\ref{MGFPTi}) is obtained.

Replacing $ W_{v_0}$ from Eq.~(\ref{W0}) in Eq.~(\ref{DWI0}), we obtain}
\begin{eqnarray}
  W_{v_i}&=& \frac{(2gm^2 - 4gm - 4im - 4m + 2im^2 - m^2 + 4)(m - 1)^g}{2E\times(m - 2)^2}\nonumber \\
  &+& \frac{ 4(m - 1)^{g -i+ 1}+m^2}{2E\times(m - 2)^2} .
\label{Wi}
\end{eqnarray}
Hence,
\begin{eqnarray}
\Sigma&=&\sum_{u\in \Omega}\pi(u)W_u\nonumber \\
  &=&\frac{mW_{v_0}+\sum_{i=1}^{g-1}m^2(m-1)^{i-1} W_{v_i}+m(m-1)^{g-1} W_{v_g}}{2E}\nonumber \\
  &=&\frac{2m(2gm^2-2m - 4gm  - m^2 + 2)(m - 1)^{2g}}{2(m - 2)^3 2E^2}\nonumber \\
  &+&\frac{  m(3m^2 + 4m - 4)(m - 1)^{g}-m^3}{2(m - 2)^3 2E^2}.
\label{WSC}
\end{eqnarray}
Therefore,
\begin{eqnarray}
\langle GFPT_0\rangle&=&E(2W_{v_0}-\Sigma)+1\nonumber \\
      &=&\frac{1}{2E}\left[(m-1)^{2g}\frac{4m(m-1)}{(m-2)^3}\right.\nonumber \\
                 & &-(m-1)^{g}\frac{m(4gm-3m+6)}{(m-2)^2}\nonumber \\
                 & &\left.-\frac{m(3m^2-8m+8)}{(m-2)^3} \right],
                \label{MGFPT0d}
\end{eqnarray}
 and Eq.~(\ref{MGFPT0}) is obtained.


\begin{thebibliography}{0}
\bibitem{Notes1}
Note: In general there are two definitions for the
   GFPT due to different ways that the starting site is selected. In the
   first, all nodes of the network have an equal probability of being
   selected as the starting site. In the second, the starting site is
   selected by mimicking the steady state, i.e., the probability that a
   node $u$ is selected as starting site is $d_u/(2E)$.  In this paper,
   we have adopted the second definition of the GFPT. On the Cayley
   trees, the mean GFPT using the first definition is obtained, but the
   mean GFPT using the second definition remains unknown.
\end{thebibliography}
\end{document}